\documentstyle[epsfig]{article}

\begin{document}
\title{Direct correlation functions and bridge functions in 
additive hard-sphere mixtures}
\author{S. B. Yuste\thanks{e-mail: santos@unex.es} and A. 
Santos\thanks{e-mail: andres@unex.es}\\ Departamento de F\'{\i}sica, 
Universidad de Extremadura,\\ E-06071 Badajoz, Spain\\ M. L\'{o}pez de 
Haro\thanks{Also Consultant at Programa de Simulaci\'{o}n Molecular del 
Instituto Mexicano del Petr\'{o}leo; e-mail: malopez@servidor.unam.mx }\\
Centro de Investigaci\'on en Energ\'{\i}a,\\
U. N. A. M.\\
Temixco, Mor.\ 62580, M{e}xico 
}
\date{\today}
\maketitle

\begin{abstract}
A method to obtain (approximate) analytical expressions for the
radial distribution functions in a multicomponent mixture of additive hard
spheres that was recently introduced is used to obtain the direct correlation functions 
and bridge functions in these systems. This method, which yields  
results practically equivalent to the
Generalized Mean Spherical Approximation and includes thermodynamic
consistency, is an alternative to the usual integral equation approaches and
requires as input only the contact values of the radial distribution
functions and the isothermal compressibility. Calculations of the 
bridge functions for a binary mixture using the
Boubl\'{\i}k-Mansoori-Carnahan-Starling-Leland equation of state are compared
to parallel results obtained from the solution of the Percus-Yevick
equation. We find that the conjecture recently proposed by Guzm\'{a}n and
del R\'{\i}o (1998, {\em Molec. Phys.}, {\bf 95}, 645) stating that 
the zeros of the  bridge functions occur
approximately at the same value of the shifted distance for all pairs of
interactions is at odds with our results. Moreover,
in the case of disparate sizes, even the Percus-Yevick  bridge
functions do not have this property. It is also found that the bridge 
functions are not necessarily non-positive.
\end{abstract}

\section{Introduction}
\label{sec.1}
Integral equation theories for the description of thermodynamic and
structural properties of liquids usually lead to qualitatively satisfactory
results. Nonetheless, in general they involve hard numerical labor as well
as criteria to formulate the closure relations that are not clearcut. 
 This is true 
even for the simplest 
and most studied systems, namely the pure hard-sphere fluid and hard-sphere 
fluid mixtures. Therefore, 
it is not surprising that many attempts at providing general features, 
symmetries, approximations 
or parametrizations of the so-called bridge functions have been reported 
in the literature 
\cite{varios,HNC,VM,MS,RY,BPGG,ZH,ZS,LLG,KBR,Rast99}. These 
 bridge functions enter in the closure relations and are defined as the sum 
 of elementary diagrams (whose precise computation is a 
formidable and rather difficult task); they account for some molecular 
spatial correlations of higher 
order than pair correlations. Of course the availability of the analytical 
results provided by the 
Percus-Yevick (PY) theory \cite{PY} in the case of the pure hard-sphere 
fluid \cite{WT} and hard-sphere fluid mixtures \cite{Lebowitz} allows one to 
determine explicitly the bridge functions in this instance, 
but they inherit the (theoretically) unpleasant lack of thermodynamic
consistency as well as the limited density range of applicability involved
in the PY approximation.

In a related context, it is worth pointing out that in the pioneering work of 
Rosenfeld and 
Ashcroft \cite{RA} it was found that an important class of pair potentials 
shared the property that their 
corresponding bridge functions were remarkably similar to each other and to 
the hard-sphere bridge 
function. This observation led to the common form of the 
reference-hypernetted chain 
theory \cite{RHNC}, considered by many to be perhaps the most accurate 
theory for the structural properties of fluids, in which the bridge 
functions of the system of interest are equated to those of a hard-sphere 
system. Thus, the search for accurate and relatively simple approximations 
for the bridge functions of a pure hard-sphere fluid and hard-sphere fluid 
mixtures has been pursued
in the last few years. Notable among the results of this pursuit are the 
empirical parametrization 
due to Malijevsk\'{y} and Lab\'{\i}k  (ML) for the hard-sphere fluid and 
its recent 
extension to binary hard-sphere mixtures \cite{Tolia}. These involve a 
careful and thorough analysis of a  large set of computer simulation data. 
Some apparent regularities of the  bridge functions in the case 
of binary mixtures (present in the PY results and in recent simulation 
data of Malijevsk\'{y} {\em et al.} \cite{MBS}) has 
been recently suggested by Guzm\'{a}n and  del R\'{\i}o \cite{GR}. Were this 
regularities to hold in general, they would allow one to simplify the ML 
parametrization and serve as a starting point to consider hard-sphere 
mixtures with three or more components. Due to the scarcity of simulation 
data for mixtures, the suggestion remains a mere conjecture up to now.

Notwithstanding the merits of all these theoretical and semiempirical
efforts, it is clear that, especially in the case of mixtures, the scarcity
of results and the relatively slow progress reflect the amount and
difficulty of the numerical work that has to be done to get them. Therefore,
one may reasonably wonder whether an alternative theoretical approach, that
at least
avoided the inherent difficulty of solving nonlinear integral equations,
would also provide information on the bridge functions of the pure
hard-sphere fluid and hard-sphere mixtures. It is the major aim of this
paper to provide an affirmative answer to the foregoing question.

In previous work \cite{YS, RFA} we have introduced a method to analytically 
derive (approximate) expressions for the radial distribution functions and 
structure factors of fluids and fluid mixtures. This method rests on a 
completely different philosophy than the one involved in integral equation 
theories and thus  is totally void of the difficulty associated with 
providing any particular closure. In the  case of pure hard-sphere fluids 
and hard-sphere mixtures and in the lowest order of approximation, it yields 
the well known PY results. Furthermore and by construction, our expressions 
in the next order of approximation, which yields  results practically 
equivalent to those of the Generalized Mean Spherical Approximation (GMSA) 
\cite{GMSA} but is much simpler to implement, also embody thermodynamic 
consistency. As shown below, by using such radial distribution functions in 
connection with the Ornstein-Zernike (OZ) equation it is rather 
straightforward to derive the direct correlation functions and, in turn, the 
bridge functions of the system,  this latter for distances greater than the 
contact distance.

The organization of the paper is the following. In Section \ref{sec.2} we 
outline the main ideas of our method to obtain the radial distribution 
functions of an $N$-component mixture of additive hard-spheres. For this 
mixture the Boubl\'{\i}k-Mansoori-Carnahan-Starling-Leland (BMCSL) 
\cite{BMCSL} equation of state 
and the Grundke-Henderson-Lee-Levesque 
(GHLL) \cite{GH,LLL} contact values of the radial distribution functions, 
which yield the BMCSL equation of state, are assumed to hold. If $N$ is set 
equal to one then the pure hard-sphere fluid case readily follows.  
Expressions for the direct correlation functions  are 
given there. Section \ref{sec.3} provides an analysis of  the bridge 
functions as well as a comparison with previous work. We 
close the paper in Section \ref{sec.4} with further discussion and some 
concluding remarks.

\section {The radial distribution functions, the direct correlation functions 
and the bridge functions of a hard-sphere mixture}
\label{sec.2}

 In this Section we outline the method to
obtain (approximate) analytical expressions for the radial distribution
functions $g_{ij}(r)$ of a  multicomponent hard-sphere mixture. 
It consists  of an extension to mixtures of the method 
previously applied to one-component systems of hard spheres, sticky 
hard spheres, and square wells \cite{YS}. For details the reader may 
refer to Ref. \cite{RFA}.

 An $N$-component mixture made of $\rho _{i}$ hard
spheres (of diameter $\sigma _{i}$) per volume unit may be characterized by 
$2N-1$  parameters 
(for instance, 
the $N-1$ molar fractions $x_{i}$,
the $N-1$ size ratios $\sigma _{i}/\sigma _{1}$ and the packing
fraction $\eta =\frac{\pi }{6}\sum_{i}\rho _{i}\sigma _{i}^{3}$) and
involves $N(N+1)/2$ radial distribution functions $g_{ij}(r)$. 

As  happens in the PY and GMSA theories, it is convenient to work in the 
Laplace space and define 
\begin{equation}
\label{Gij(s)}
G_{ij}(s)=\int_{0}^{\infty}dr\,e^{-sr}rg_{ij}(r).
\end{equation} 
There are two basic requirements that $G_{ij}(s)$ must satisfy. 
First, since $g_{ij}(r)=0$ for $r<\sigma _{ij}$, 
with $\sigma_{ij}=
\left( \sigma _{i}\ +\sigma _{j}\right) /2$, and 
$g_{ij}(\sigma_{ij}^{+})=\mbox{finite}$, this implies that 
(i) $\lim_{s\rightarrow \infty}s\,e^{s\sigma _{ij}}G_{ij}(s)=\mbox{finite}$. 
Second, the isothermal compressibility $\kappa _{T}=\mbox{finite}$, so that 
(ii) $\lim_{s\rightarrow 0}[G_{ij}(s)-s^{-2}]=\mbox{finite}$. 
The approximation we will use consists of assuming the
following functional form: 
\begin{equation}
G_{ij}(s)=\frac{e^{-s\sigma _{ij}}}{2\pi s^{2}}\sum_{k}L_{ik}(s){[(1+\alpha
s){\sf I}-{\sf A}(s)]^{-1}}_{kj},  \label{G(s)}
\end{equation}
where ${\sf I}$ is the $N\times N$ unit matrix,
\begin{equation}
\label{L(s)}
L_{ij}(s)=L_{ij}^{(0)}+L_{ij}^{(1)}s+L_{ij}^{(2)}s^{2}
\end{equation}
 and 
\begin{equation}
\label{A(s)}
A_{ij}(s)=\rho _{i}\sum_{n=0}^{2}\varphi _{n}(s\sigma _{i})\sigma
_{i}^{n+1}L_{ij}^{(2-n)},
\end{equation}
 with 
\begin{equation}
\label{phi}
\varphi _{n}(x)\equiv 
x^{-(n+1)}\left[\sum_{m=0}^{n}\frac{(-x)^{m}}{m!}-e^{-x}\right].
\end{equation}

Condition (i) is verified by
construction. On the other hand, condition (ii) yields two {\em linear\/}
sets of $N^{2}$ equations each, whose solution is straightforward, namely 
\begin{equation}
\label{L0}
L_{ij}^{(0)}=\lambda +\lambda ^{\prime }\sigma _{j}+2\lambda ^
{\prime}\alpha -\lambda \sum_{k}\rho _{k}\sigma _{k}L_{kj}^{(2)},
\end{equation}
\begin{equation}
\label{L1} 
L_{ij}^{(1)}=\lambda \sigma _{ij}+\frac{\lambda ^{\prime }}{2}\sigma
_{i}\sigma _{j}+(\lambda +\lambda ^{\prime }\sigma _{i})\alpha -\frac{
\lambda }{2}\sigma _{i}\sum_{k}\rho _{k}\sigma _{k}L_{kj}^{(2)},
\end{equation}
where $\lambda \equiv 2\pi /(1-\eta )$ and $\lambda ^{\prime }\equiv
(\lambda /2)^{2}\sum_{k}\rho _{k}\sigma _{k}^{2}$. 

The parameters 
$L_{ij}^{(2)}$ and $\alpha $ (which play a role similar to that of the
parameters $K_{ij}$ and $z$ in the GMSA) are arbitrary, so that conditions 
(i) and (ii) are satisfied 
regardless of their 
choice. In particular, if one chooses $L_{ij}^{(2)}=\alpha =0$, our 
approximation coincides with 
the PY solution. 
If, on the other hand, we fix given values for 
$g_{ij}(\sigma_{ij}^{+})$, we get the relationship 
\begin{equation}
\label{L2}
L_{ij}^{(2)}=2\pi \alpha \sigma_{ij}g_{ij}(\sigma _{ij}^{+}).
\end{equation}
 Thus, only $\alpha $ remains to be determined. Finally, if we fix 
 $\kappa _{T}$, we obtain a closed 
equation
for $\alpha $ of degree $2N$. It is worth pointing out that in the particular 
case of a pure hard-sphere fluid ($N=1$) one gets a quadratic algebraic 
equation for $\alpha$, while for a
binary mixture 
($N=2$) the {\em explicit\/} knowledge of $G_{ij}(s)$ only requires to 
solve a quartic equation, 
which also has an {\em analytical\/} solution. 
A natural choice to close the scheme, which we will of course consider
 in this paper, is to
 take the GHLL values  \cite{GH,LLL} of 
$g_{ij}(\sigma _{ij}^{+})$,  
 as 
well as the corresponding BMCSL  \cite{BMCSL} isothermal 
compressibility $\kappa _{T}$.
 But other possibilities are available
and one of them will also be addressed later on. 
Once $G_{ij}(s)$ has been determined, inverse Laplace transformation
directly yields $rg_{ij}(r)$, while the Fourier transforms 
$\widetilde{h}_{ij}(q)$ of the total correlation functions $h_{ij}(r)$ 
readily follow from the relation 
\begin{eqnarray}
\label{H(q)}
\widetilde{h} _{ij}(q)&\equiv&\int d{\bf r}\,\exp(\imath {\bf q}\cdot{\bf 
r})h_{ij}(r)\nonumber\\
&=&-2\pi
	\left. 
\frac{G_{ij}(s)-G_{ij}(-s)}{s}\right| _{s=\imath q},
\end{eqnarray}
where $\imath$ is the imaginary unit.
In Fourier space and introducing the quantities 
$\widehat{H}_{ij}(q)=
\sqrt{\rho _{i}\rho _{j}}\widetilde{h}_{ij}(q)$  and 
$\widehat{C}_{ij}(q)=\sqrt{\rho_{i}\rho_{j}}\,\widetilde{c}_{ij}(q)$ the 
OZ equation reads 
\begin{equation}
\label{C(q)}
\widehat{\sf C}(q)=
\widehat{\sf H}(q)\cdot[{\sf I}+\widehat{\sf H}(q)]^{-1} ,
\end{equation}
so that after replacement of $\widehat{\sf H}(q)$ and
subsequent inverse Fourier transformation it is straightforward
to get $c_{ij}(r)$.
The result  gives $c_{ij}(r)$ for $r>\sigma_{ij}$ as the superposition 
of $N$ Yukawas (see the Appendix):
\begin{equation}
 c_{ij}(r)= \sum_{\ell=1}^N 
 \frac{K_{ij}^{(\ell)}}{r}\exp\left[-z_\ell(r-\sigma_{ij})\right],
\label{c(r)}
\end{equation}
where $q=\pm \imath z_\ell$ with $\ell=1,\ldots,N$ are the zeros of 
$\det\left[{\sf I}+\widehat{\sf H}(q)\right]$ and the
 amplitudes $K_{ij}^{(\ell)}$ are obtained by applying the residue theorem as
\begin{equation}
 K_{ij}^{(\ell)}=\frac{\imath z_\ell}{2\pi} e^{-z_\ell \sigma_{ij}} 
 \lim_{q\rightarrow \imath z_\ell} \widetilde{c}_{ij}(q)(q-\imath z_\ell).
\label{kij}
\end{equation}

 Finally, we note that the bridge functions $B_{ij}(r)$
for $r>\sigma_{ij}$ are linked to $g_{ij}(r)$ and $c_{ij}(r)$ through
\begin{equation}
 B_{ij}(r)= \ln g_{ij}(r)-g_{ij}(r)+c_{ij}(r)+1.
\label{b(r)}
\end{equation}
 Equations (\ref{c(r)}) and (\ref{b(r)}), after replacement of the results 
 for $g_{ij}(r)$, will be used
below to investigate some properties of the bridge functions $B_{ij}(r)$ 
in pure hard-sphere fluids and hard-sphere mixtures.

\section {Comparison with other results}
\label{sec.3}

\begin{figure}[h]
\begin{center}
\parbox{0.5\textwidth}{\epsfxsize=\hsize\epsfbox{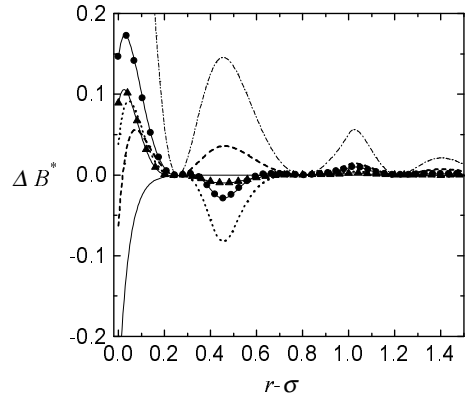}}
\caption{
The difference $\Delta B^{*} (r)$ as a function of the shifted distance
$r-\sigma$ for a simple hard-sphere fluid with $\eta=0.49$ and $\sigma=1$, 
according to various closures: HNC (dash-dotted line), PY (solid
line), RY (dashed line), MS (dotted line), VM (solid line with  triangles) 
and BPGG (solid line with  circles).
\label{fig1}}
\end{center}
\end{figure}

We begin with the pure hard-sphere fluid, that is, we now 
consider the case when $N=1$. For this system a variety of closures to the 
OZ equation are available \cite{HNC,VM,MS,RY,BPGG,ZH}, 
for instance
\begin{equation}
\label{CPY}
B^{\mbox{\scriptsize PY}}(r)=\ln [1+\gamma(r)]-\gamma(r)\; ,
\end{equation}
\begin{equation}
\label{CHNC}
B^{\mbox{\scriptsize HNC}}(r)=0\; ,
\end{equation}
\begin{equation}
\label{CVM}
B^{\mbox{\scriptsize VM}}(r)=-\frac{\left[\gamma(r)\right]^2}{2
\left\{1+a_{1} 
\gamma(r)\right\}}\; , \end{equation}
\begin{equation}
\label{CMS}
B^{\mbox{\scriptsize 
MS}}(r)=\left[1+2\gamma(r)\right]^{1/2}-\gamma(r)-1\; ,
\end{equation}
\begin{equation}
\label{CRY} 
B^{\mbox{\scriptsize RY}}(r)=\ln\left\{1+\frac{\exp\left\{[1-\exp(-a_{2}  
r)]\gamma(r)\right\}-1}{1-\exp(-a_{2}  r)} \right\}-\gamma(r)\; 
, \end{equation}
\begin{equation}
\label{CBPGG}
B^{\mbox{\scriptsize BPGG}}(r)=\left[1+a_{3} 
\gamma(r)\right]^{1/a_{3} }-\gamma(r)-1\; , 
\end{equation}
where $\gamma(r)\equiv g(r)-c(r)-1$ and the labels HNC, VM,  MS, RY  and 
BPGG denote the hypernetted-chain \cite{HNC}, Verlet 
modified \cite{VM}, Martynov-Sarkisov \cite{MS}, Rogers-Young \cite{RY}    
and Ballone-Pastore-Galli-Gazzillo \cite{BPGG} closures, respectively,  and 
the  adjustable parameters $a_{i} (i=1,2,3)$ have been estimated to take the 
values $a_{2}=0.160$ \cite{RY}, $a_{1}=4/5$ and $a_{3}=15/8$ \cite{LLG}. It 
is at this point interesting to analyze to what extent the bridge functions 
calculated upon substitution in such closures of our expressions for the 
radial distribution function and direct correlation function differ or are 
compatible with the actual bridge function computed using equation 
(\ref{b(r)}). 
 To this end, we introduce the quantity $\Delta B^{*} (r) \equiv B^{*}(r) - 
B(r)$, where the asterisk refers to a given label. Notice that in particular 
$\Delta B^{\mbox{\scriptsize HNC}} (r) = -B(r)$ and $\Delta 
B^{\mbox{\scriptsize PY}} (r) =\ln[1-c(r)/g(r)]$. In figure \ref{fig1} we 
display the behavior of $\Delta B^{*} (r)$ as a function of the shifted 
distance $r-\sigma$ for a packing fraction $\eta=0.49$, i.e. close to the 
freezing transition. Clearly the main discrepancies between both types of 
calculation show up near the contact point, but one could reasonably argue 
that the RY, the MS and the VM closures are rather compatible with the 
result of the present approach in that region. 

\begin{figure}[h]
\begin{center}
\parbox{0.5\textwidth}{\epsfxsize=\hsize\epsfbox{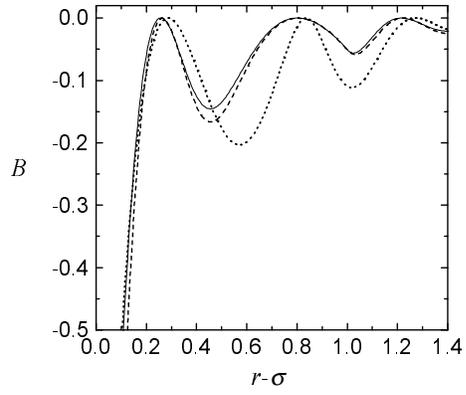}}
\caption{
Bridge function $B(r)$  versus $r-\sigma$ for a simple hard-sphere fluid 
with $\eta=0.49$ and $\sigma=1$.
Solid line: present method; dashed line: PY results; dotted line: 
ML parametrization.
}
\label{fig2}
\end{center}
\end{figure}

Figure \ref{fig2} displays the results for the bridge function as 
obtained with the PY theory, the parametrization of Malijevsk\'{y} and 
Lab\'{\i}k (ML)
\cite{Tolia} and our 
formulation, again for the packing fraction $\eta =0.49$. As clearly seen 
in the figure, the discrepancy 
between the results of the
 parametrization and both ours and those of the PY theory is rather 
 significant.
Also, although not perceptible in the scale of the figure, we note that in 
our case the first maximum of the bridge function 
attains a positive value, 
whereas both the ML parametrization and the PY 
theory always lead to non-positive values for the bridge function. 
We will come back to this point later on.

\begin{figure}[h]
\begin{center}
\parbox{0.5\textwidth}{\epsfxsize=\hsize\epsfbox{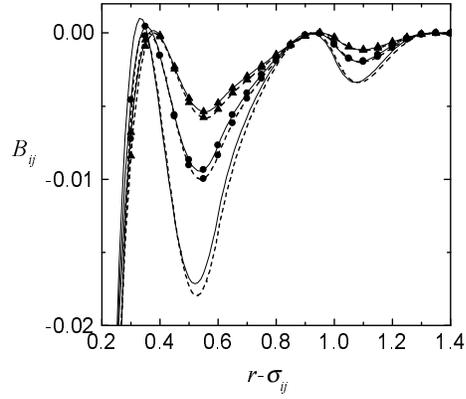}}
\caption{
 Bridge functions $B_{ij}(r)$  versus $r-\sigma_{ij}$ for an equimolar 
 binary mixture of hard spheres with $\eta=0.35$  and diameters  
 $\sigma_2=0.6$ and $\sigma_1=1$. 
Solid lines: present method; dashed lines: PY results. The curves for 
$B_{11}(r)$ contain no symbols, those for $B_{12}(r)$ contain 
circles and the ones for $B_{22}(r)$ contain  triangles.
}
\label{fig3}
\end{center}
\end{figure}

\begin{figure}[h]
\begin{center}
\parbox{0.5\textwidth}{\epsfxsize=\hsize\epsfbox{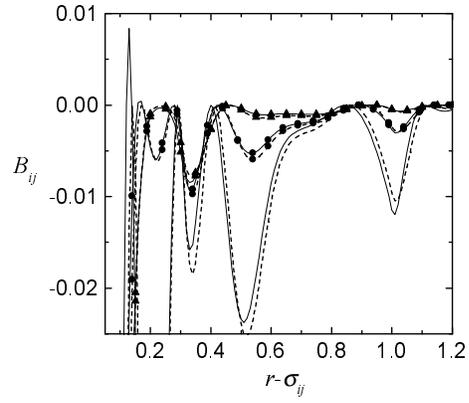}}
\caption{
Bridge functions $B_{ij}(r)$  versus $r-\sigma_{ij}$ for a binary mixture 
of hard spheres with $\eta=0.49$, molar fraction $x_1=1/16$  
and diameters $\sigma_2=0.3$ and $\sigma_1=1$. 
The code for the different curves is as in figure \protect\ref{fig3}.
}
\label{fig4}
\end{center}
\end{figure}

We now turn to binary mixtures. In figures \ref{fig3} and \ref{fig4} results 
for the different  bridge functions are shown for two cases. In figure 
\ref{fig3},  which corresponds to the case examined by Guzm\'{a}n and del 
R\'{\i}o \cite{GR}, it is an equimolar mixture with $\eta=0.35$ and 
$\sigma_{2}/\sigma_{1}=0.6$;  the second mixture is 
defined by $\eta=0.49$, $x_{1}=1/16$, $x_{2}=15/16$ and 
$\sigma_{2}/\sigma_{1}=0.3$. We have also included in these figures the 
results obtained with the PY approximation. The differences 
between the results of both approaches are particularly
important in the first maximum of the bridge functions and once more in our 
case we get some intervals of positive 
values. These 
figures also illustrate the fact that the phase-shift symmetry that was 
recently conjectured by Guzm\'{a}n and del R\'{\i}o \cite{GR} to hold  
on the basis of the behavior observed in figure \ref{fig3}, is not even 
present in the PY theory, particularly 
at short distances, 
and this becomes more evident 
as the disparity in size ratio is increased, as shown in figure \ref{fig4}. 

\begin{figure}[h]
\begin{center}
\parbox{0.5\textwidth}{\epsfxsize=\hsize\epsfbox{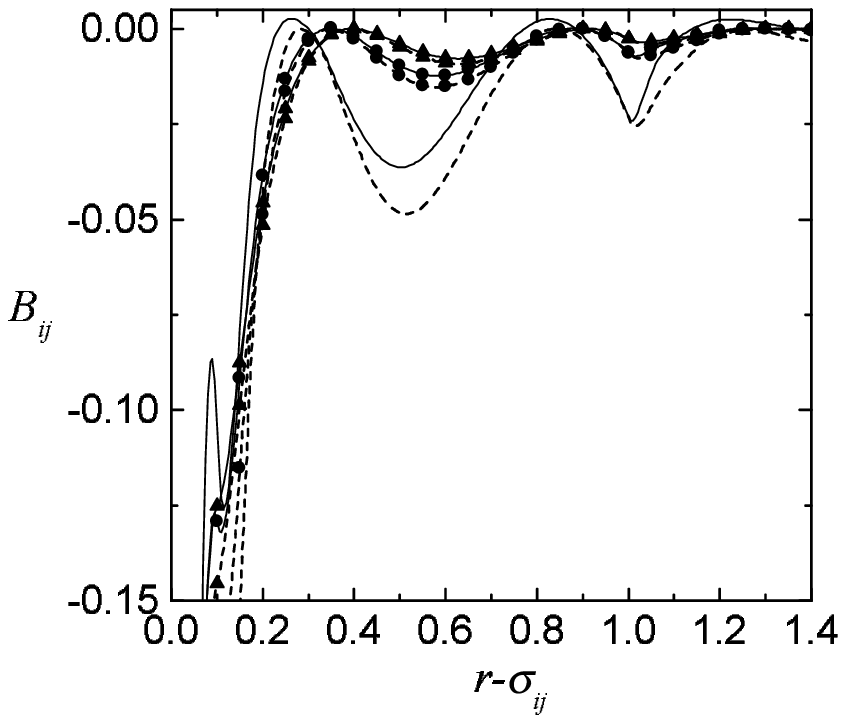}}
\caption{
Bridge functions $B_{ij}(r)$  versus $r-\sigma_{ij}$ for a ternary mixture 
of hard spheres with $\eta=0.49$, molar fractions $x_1=x_2=1/102$ and 
diameters $\sigma_3=0.1$, $\sigma_2=0.3$ and 
$\sigma_1=1$. Solid lines: present method; dashed lines: PY results. The 
curves for $B_{11}(r)$ contain no symbols, those for $B_{12}(r)$ contain 
 circles  and the ones for $B_{13}(r)$ contain  triangles.
}
\label{fig5}
\end{center}
\end{figure}

\begin{figure}[h]
\begin{center}
\parbox{0.5\textwidth}{\epsfxsize=\hsize\epsfbox{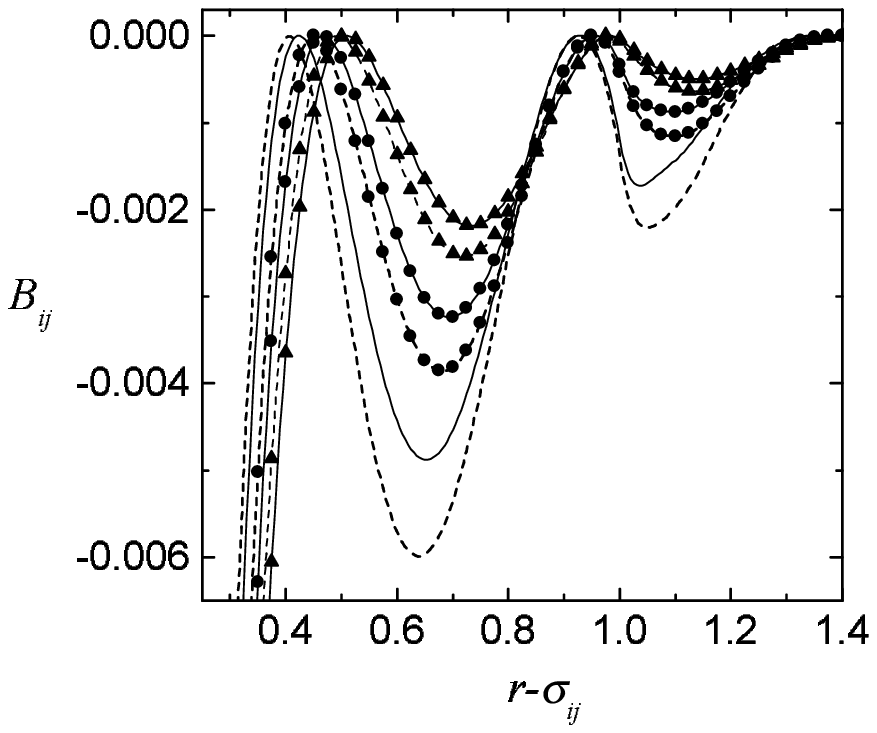}}
\caption{
Bridge functions $B_{ij}(r)$  versus $r-\sigma_{ij}$ for a ternary mixture
 of hard spheres with $\eta=0.49$, molar fractions $x_1=x_2=1/102$, and 
diameters $\sigma_3=0.1$, $\sigma_2=0.3$ and 
$\sigma_1=1$.
Solid lines: present method; dashed lines: PY results. The curves for 
$B_{22}(r)$ contain no symbols, those for $B_{23}(r)$ contain 
circles  and the ones for $B_{33}(r)$ contain  triangles.
}
\label{fig6}
\end{center}
\end{figure}

 As another illustration, in figures \ref{fig5}  and \ref{fig6} we display 
 results for a
ternary mixture where $\eta =0.49$,  $x_{1}=x_{2}=1/102$, $x_{3}=100/102$, 
$\sigma
_{2}/\sigma _{1}=0.3$ and $\sigma _{3}/\sigma _{1}=0.1$. Apart from
exhibiting a more complicated structure than in the case of 
binary mixtures ---notice, for example, 
the existence 
of a negative first maximum for $B_{11}$ and $B_{12}$---, again 
an important difference 
between our results
 and those of the PY theory is that we may get positive values for the bridge 
 functions in some regions.

\begin{figure}[h]
\begin{center}
\parbox{0.5\textwidth}{\epsfxsize=\hsize\epsfbox{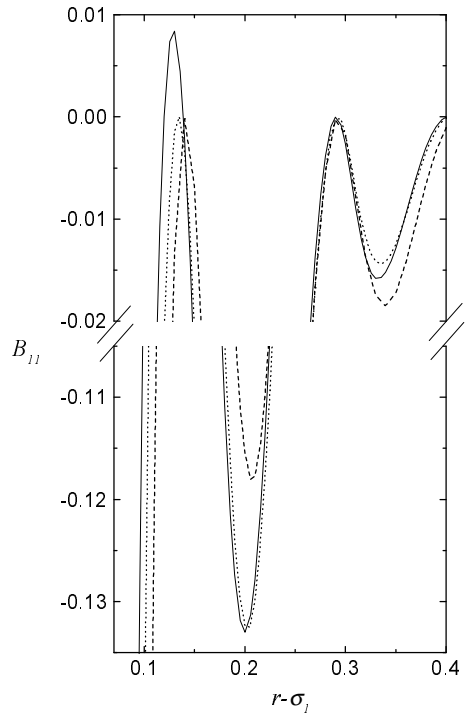}}
\caption{
Bridge function $B_{11}(r)$  versus $r-\sigma_{1}$ for a binary mixture 
of hard spheres with $\eta=0.49$, molar fraction $x_1=1/16$  
and diameters $\sigma_2=0.3$ and $\sigma_1=1$. 
Solid line: present method using the GHLL contact values
of $g_{ij}(\sigma_{ij}^{+})$ and the compressibility $\kappa _{T}$ derived 
from the BMCSL equation of state; dotted line: present method using the 
contact values
of $g_{ij}(\sigma_{ij}^{+})$ obtained by extrapolation of the 
simulation data in 
 Ref.\ \protect\cite{MBS} (namely, $g_{11}=10.23$, $g_{12}=4.69$ and 
 $g_{22}=3.57$) and the compressibility $\kappa _{T}$ derived from 
the equation of state eCS
proposed
in Ref.\ \protect\cite{Nos}; dashed line: PY results.
}
\label{fig7}
\end{center}
\end{figure}

Thus far we have only considered the GHLL prescription for the contact 
values $g_{ij}(\sigma _{ij}^{+})$  and the 
isothermal compressibility  $\kappa _{T}$ derived from the BMCSL equation of 
state as the input in our method. One may reasonably wonder whether the use 
of different values for $g_{ij}(\sigma _{ij}^{+})$ and/or $\kappa _{T}$ 
would also yield similar results.
 In order to assess the importance of other reasonable choices, we 
have made calculations using the contact values $g_{ij}(\sigma _{ij}^{+})$ 
obtained by extrapolation of simulation results \cite{MBS} and
the value of $\kappa_{T}$  derived from an
equation of state for mixtures (eCS) recently proposed by us 
\cite{Nos}. 
To 
illustrate the results one gets with these choices, in figure \ref{fig7} we 
compare the various calculations of the function $B_{11}(r)$ for the case 
considered earlier in figure \ref{fig4}, namely the binary mixture defined 
by $\eta=0.49$, $x_{1}=1/16$, $x_{2}=15/16$ and $\sigma_{2}/\sigma_{1}=0.3$. 
Except in the region up to  the first maximum and near the second minimum, 
the curves obtained with
either the BMCSL equation of state and with the second choice (eCS) using 
our procedure
are practically indistinguishable. Nevertheless, although the
first maximum with the eCS choice is still positive, its amplitude is much 
smaller than the one using the BMCSL equation of state, up to
a point that the positive character can be hardly ascertained in the scale of the 
figure. It is also worth pointing out that the value of $\alpha $ is more sensitive 
to the choice of $\kappa _{T}$ than to the values of $g_{ij}(\sigma 
_{ij}^{+})$. In fact, for 
this mixture one gets $\alpha=0.0189$ using the BMCSL equation of 
state while 
$\alpha=0.0118$ using our proposal (eCS) for the equation of state.
A smaller value of $\alpha$ means that $c_{11}(r)$ goes to zero more 
rapidly which in turn implies a much smaller (but still positive) value for 
the first maximum of $B_{11}(r)$ in the latter case.    

\section{Discussion}
\label{sec.4}
The points arising from the results of the previous sections deserve 
further elaboration. To begin with, it is fair to say that our approach 
leads to a simple and clearcut procedure to determine both the bridge 
functions and the direct correlation functions in a multicomponent 
hard-sphere mixture, requiring as the {\it only input} the contact values of 
the radial distribution functions that specify the actual equation of state 
of the mixture. It should be pointed out that while the procedure is capable 
of yielding the values of the direct correlation functions for all 
distances, including those inside the hard cores, in the case of the bridge 
functions it is limited to distances greater than the contact distance. This 
is due to the fact that our method does not deal neither with closures nor 
with the cavity functions. Nevertheless, this restriction may be disposed 
of, at least for the case of the pure hard-sphere fluid, by considering 
approximate analytical forms of the cavity function that are available in 
the literature \cite{ZS,GH,Speedy}. In this connection, we should mention 
that the form of the cavity function derived by Zhou and Stell \cite{ZS} has 
been shown to be compatible with our $g(r)$ in the sense that it yields the 
same values for both $g(\sigma^+)$ and $\left.dg(r)/dr\right|_{r=\sigma^+}$ 
\cite{Miguel}. 
 If the available simulation results for the radial distribution functions 
for mixtures are scarce, those for the bridge functions are to our knowledge 
nonexistent. In the absence of such data to compare with the results we have 
presented, it would be of course premature to reach definite conclusions. 
One could argue that the accuracy of the bridge functions might
be estimated indirectly by comparing the radial distribution functions
calculated using the OZ equation with a given closure and simulation
results. In the present approach, however, this is unnecessary since we have
explicit (analytical) expressions for the radial distribution functions from
the very beginning, and these have been already compared to simulation
results both for the one-component and two-component hard-sphere systems in
Refs.\ \cite{YS} and \cite{RFA}. 

It is clear that a key difference between 
our results for the bridge
functions and most of those previously reported, is the fact that in our 
case
 these functions may attain both {\it positive 
and negative} values. In connection with this issue one cannot overlook the 
fact that it has often been assumed that the bridge functions should be 
nonpositive. This is certainly the case in the PY theory and various 
parametrizations and approximations have included such an assumption. 
Nevertheless, as Rast {\em et al.} \cite{Rast99} have 
recently pointed out, there seems to be no rigorous reason or argument 
stating that it should be so. 
In fact, any theory that leads to a positive value of the direct correlation 
function $c(r)$ at a distance where $g(r)=1$ will produce a positive $B(r)$ 
at that distance. For instance, taking the Monte Carlo data for $c(r)$ 
obtained by Groot {\em et al.} \cite{Groot} and those of $g(r)$ given by 
Barker and Henderson \cite{BH72}, one finds that $B(r)\simeq 0.2$ for 
$r\simeq 1.85\sigma$ and $\eta=0.445$.
 In further support of the likely correctness of our 
results, one can invoke the fact that in our case thermodynamic consistency 
is an ingredient of the formulation while for instance the PY theory does 
not share this property. For the pure hard-sphere fluid, the reasonable 
compatibility between our bridge function and the one computed using in 
particular the RY closure (which was originally proposed to achieve 
thermodynamic consistency)  is very satisfactory in 
this respect. Further, the fact that our $g_{ij}(r)$ are in better agreement 
with simulation results than those of the PY theory particularly in the 
region around the contact point \cite{RFA} also favors the present approach.

Concerning the ML parametrization for mixtures and the apparent regularity 
of the shifted bridge functions that was conjectured to hold in general in 
Ref. \cite{GR}, we can only add that unfortunately it does not do so. 
Indeed, it would have been rather remarkable that the relatively simple 
forms proposed for the bridge functions would have been able to capture the 
rich and varied behaviors that one would expect from the number of 
parameters involved in the description of mixtures.
 
Finally, we want to point out that due to the similarity of the
bridge functions corresponding to different potentials and those of 
hard-spheres  \cite{RA} ---in fact the {\it universality\/} of the
hard-sphere bridge functional has been recently shown to be very reliable 
\cite{KBR}--- these results are not only relevant for hard-sphere mixtures,
 but they may also prove useful in connection with the integral equation 
 approach in liquid theory for mixtures with 
other interaction potentials.

Fruitful discussions with F. del R\'{\i}o, who suggested this problem to 
us, are gratefully acknowledged.
One of us (M.L.H.) wants to thank the Universidad de Extremadura at 
Badajoz, where the draft of the paper was prepared, for its hospitality. He 
also acknowledges the partial support of DGAPA-UNAM under Project IN-117798.
The work of two of us (S.B.Y. and A.S.) has been partially supported by 
DGES (Spain) through grant No. PB97-1501 and by the Junta de 
Extremadura--Fondo Social Europeo through grant No.\ IPR98C019.

\appendix
\section{}
{}From the Fourier transform $\widetilde{c}_{ij}(q)$ one can get the direct 
correlation function in real space as
\begin{eqnarray}
\label{A1}
{c}_{ij}(r)&=&\frac{1}{(2\pi)^3}\int d{\bf q} \, \exp(-\imath {\bf 
q}\cdot{\bf r})\widetilde{c}_{ij}(q)\nonumber\\
&=&\frac{1}{4\pi^2\imath r}\int_{-\infty}^\infty dq\, q e^{\imath q 
r}\widetilde{c}_{ij}(q).
\end{eqnarray}
It is now convenient to see $q$ as a complex variable. Thus, if 
$r>\sigma_{ij}$, it is possible to take a contour integration around the 
upper half plane in equation (\ref{A1}). According to Eq.\ (\ref{C(q)}),  
the functions $\widetilde{c}_{ij}(q)$ have a {\em common\/} set of poles, 
namely the zeros 
 of $D(q)\equiv\det\left[{\sf I}+\widehat{\sf H}(q)\right]$. A careful 
inspection of the results obtained from our method shows that the zeros of 
$D(q)$ are the roots of a polynomial in $q^2$ of degree $N$. More 
specifically, the zeros of $D(q)$ are $q=\pm\imath z_{\ell}$, where the 
$z_\ell$ ($\ell=1,\ldots,N$) are positive  real numbers.

Application of the residue theorem yields
\begin{equation}
\label{A2}
c_{ij}(r)=\frac{\imath}{2\pi r}\sum_{\ell=1}^N z_{\ell} e^{-z_{\ell} r} 
\lim_{q\to\imath z_{\ell}} \widetilde{c}_{ij}(q) (q-\imath z_{\ell}),\quad 
r>\sigma_{ij},
\end{equation}
which is equivalent to Eqs.\ (\ref{c(r)}) and (\ref{kij}). To be more 
explicit, let us rewrite Eq.\ (\ref{C(q)}) as
\begin{equation}
\label{A3}
\widehat{\sf C}(q)=
{\sf I}-{\sf F}^{-1}(q) ,
\end{equation}
where ${\sf F}(q)\equiv{\sf I}+\widehat{\sf H}(q)$. Therefore, Eq.\ 
(\ref{kij}) is equivalent to
\begin{equation}
 K_{ij}^{(\ell)}=-\frac{e^{-z_\ell \sigma_{ij}}}{4\pi\sqrt{\rho_i\rho_j}}  
 \lim_{q\rightarrow \imath z_\ell} \left[{\sf 
 F}^{-1}(q)\right]_{ij}(q^2+ z_\ell^2).
\label{A4}
\end{equation}

\newpage

\newpage

\newpage

\newpage

\newpage

\newpage

\newpage

\end{document}